\documentclass[%
 reprint,
 amsmath,amssymb,
]{revtex4-2}
\usepackage{graphicx}
\usepackage{dcolumn}
\usepackage{bm}

\begin{document}

\preprint{APS/123-QED}

\title{Exact and fast solution of ultrarelativistic multibunch instabilities in a constant gradient accelerator}

\author{Muhammad Shumail}
 \email{shumail@alumni.stanford.edu}

\affiliation{%
 SLAC National Accelerator Laboratory, Menlo Park, California 94025, USA\\
}%

\date{\today}

\begin{abstract}
The electromagnetic wakefields excited by the bunches of charged particles in the particle accelerators may cause instability in the longitudinal and transverse motion of these bunches leading to the beam loss. These multibunch effects need to be analyzed carefully in an accelerator design to ensure the stability of the charged particle beams. This paper presents an analytical method describing the multibunch transverse instabilities for ultrarelativistic electron bunches whose energy changes linearly through the accelerating section of interest. Fast algorithmic implementations of this analytical method are presented for the general as well as the specific cases and this method is demonstrated with an example. This tool can also be applied to study single-bunch beam breakup. 
\end{abstract}

\maketitle


\section{\label{sec:intro} Introduction }
In the microwave particle accelerators the multibunch beams of charged particles excite electromagnetic wakefields ~\cite{Bane:1984, Novo:1988, Wilson:1989, Wangler:2008a, Wiedemann:2007} in the accelerating cavities and other beam-line components. These wakefields act back upon the bunches. The multibunch beam breakup is the phenomenon of unstable deflection of the bunches by the transverse wakefields~\cite{Wangler:2008a, Wiedemann:2007, Loew:1969, Chao:1993}. Such collective effects are of major concern in high current and high efficiency accelerators~\cite{NLC:1996, Braun:2008, Galayda:2014, Shumail:2018}, therefore, a careful analysis of the multibunch instabilities is essential in the design of such accelerators.

Generally, the dominant transverse wakefield generated by a bunch is of the dipole kind with its magnitude proportional to the off-axis displacement of this bunch. Thus, a certain bunch in a particle beam experiences the cumulative effect of the wakefields induced by all the leading bunches. This could potentially lead to the unstable deflection of the following bunches and, consequently, the beam loss in the accelerator walls. For long linear accelerators (linacs), it is usual to employ a magnetic focusing system to confine the bunch motion close to the nominal trajectory~\cite{Courant:1958, Wangler:2008b}. We have already presented an exact solution of multibunch instabilities for the case of bunches coasting with constant energy through a focusing system~\cite{Shumail:2019}. However, the overhead of a focusing system is generally avoided in the compact medical and industrial linacs where the beam is accelerated to the required energy in few meters. Thus, in these linacs the evolution of the transverse positions of the bunches needs to be determined by taking into account the effects of transverse wakefields and the longitudinal acceleration only.

The reader is referred to Ref.~\cite{Shumail:2019} for a brief review of various studies  Ref.~\cite{Schulte:2009, Panofsky:1968, Gluckstern:1985, Hoffstaetter:2004, Bohn:1991, Mosnier:1993, Ferrario:2017, Chao:1980, Bane:2000, Thompson:1990, Colombant:1988} done for mutibunch instabilities.

For ultrarelativistic bunches, as is generally the case in the electron accelerators, the transverse deflecting wakefield produced by a certain bunch can only affect the following bunches due to causality. Thus, the evolution of the transverse position of the $k$th bunch is dependent on the transverse positions of the leading bunches: $1^{\rm st}, 2^{\rm nd},...,(k-1){\rm th}$. This unidirectional effect has been described in terms of a lower triangular matrix~\cite{Schulte:2009, Shumail:2019}. In this paper, we have applied a similar approach to obtain an exact solution of transverse multibunch instabilities for the case of uniformly accelerated (with constant gradient) ultrarelativistic bunches in the absence of a focusing system as is usually the case in compact linacs used for medical, industrial, border security, and environmental applications.

In our analysis, we first consider the general case where each bunch can have different charge and the longitudinal separation between the consecutive bunches is also arbitrary. Then, we consider the specific but more usual case of the same charge for all bunches and the same interbunch longitudinal separation.  For both the general and the specific case, the rigorous analytical solution is followed an efficient algorthimic implementation that also summarizes the overall method. Finally, we present a realistic example to demonstrate the application of the analytical tool developed here. We show that it generally takes this tool only a few seconds on a typical computer to calculate the evolution of the transverse coordinates of tens of thousands of bunches in a bunch train that are accelerated through a constant gradient accelerator.

\section{\label{sec:general} General case }

Consider the point-like bunches of charge particles moving along the axis ($z$-direction) of a linac. The longitudinal and transverse positions of these bunches are denoted as $z_k$, and $x_k$, respectively, while their charge is denoted as $c_k$. Here, $k$ denotes the index of a particular bunch and $k=1$ corresponds to the leading bunch. We assume that all bunches are uniformly accelerated (or decelerated) from an initial energy $E(0)$ (including the rest mass energy) with a constant energy gradient $G$ along the linac. Thus, the energy of these bunches at a certain longitudinal position $z$ is given as follows,
\begin{equation}\label{eq:01}
E(z)=E(0) + G\ z.
\end{equation}

In the absence of any focusing element, the equation of motion for the transverse coordinate of a particle of charge $e$ in the $k$th bunch is given as (compare with Eq.~(1) in Ref.~\cite{Shumail:2019}),
\begin{equation} \label{eq:02}
\frac{d}{dz}\left( E(z)\frac{d}{dz}x_k(z) \right)= e \sum_{j=1}^{k-1}c_j W(z_{j}-z_{k})x_j(z).
\end{equation}

Here, $z_{j}-z_{k}$ indicates the relative longitudinal distance by which the $k$th bunch lags the $j$th bunch. The transverse wakefield function $W(z_{j}-z_{k})$ is the force per unit charge on the particles of $k$th bunch (witness) due to the induced electromagnetic fields per unit charge and per unit offset of the $j$th bunch (source). Thus, the units of this function are N/C\textsuperscript{2}/m or, more commonly, V/pC/mm/m. The corresponding equation for the other transverse coordinate $y_k$ is similar to Eq.~(\ref{eq:02}) and can be solved independently. Before proceeding, let us define some dimensionless quantities as follows,
\begin{subequations} \label{eq:03}
\begin{eqnarray}
s \equiv ~&& \frac{G}{E(0)} z, \label{eq:s} \\
A_{k,j} \equiv ~&&\left\{ \begin{array}{ll}
            0, &j > k-1, \\
             \frac{e E(0)}{G^2}c_j W(z_{j}-z_{k}), &\text{otherwise.}
        \end{array}
 \right.  \label{eq:Akj}
\end{eqnarray}
\end{subequations}

Using these definitions, Eq.~(\ref{eq:02}) can be simply written as follows,
\begin{equation}
(1+s) \frac{d^2}{{ds}^2}x_k(s)+\frac{d}{ds}x_k(s)=\sum_{j=1}^{k-1}A_{k,j} x_j(s), \nonumber
\end{equation}
\begin{equation}
(1+s) \textbf{\textit{x}}''(s)+\textbf{\textit{x}}'(s) = \textbf{\textit{A}} \cdot \textbf{\textit{x}}(s). \label{eq:04}
\end{equation}

In Eq.~(\ref{eq:04}), we have employed the matrix notation (as in Ref.~\cite{Schulte:2009, Shumail:2019}). Here,  $n$ being the total number of bunches that are being studied, $x_k (s)$ and $A_{k,j}$ are the elements of the $n \times 1$ vector $ \textbf{\textit{x}}(s)$ and the $n \times n$ normalized wakefield matrix $ \textbf{\textit{A}}$, respectively. The prime symbol $'$ in this paper indicates the derivative with respect to $s$. All the vectors in this analysis will be of size $n \times 1$ and denoted by the lowercase bold letters. Similarly all the matrices here will be of size $n\times n$ and denoted by the uppercase bold letters.

In order to find the solution of Eq.~(\ref{eq:04}) let us start by analyzing the structure of matrix $\textbf{\textit{A}}$ which is a consequence of causality. We notice that $\textbf{\textit{A}}$ is a hollow (diagonal elements being 0), lower triangular matrix. This rendes $\textbf{\textit{A}}$ with some useful properties. For example, all the elements of first $j$ rows in the matrix $\textbf{\textit{A}}^j\equiv \underbrace{\textbf{\textit{A}}\cdot \textbf{\textit{A}} \cdot \cdot \cdot \textbf{\textit{A}}}_{j~\text{terms}}$ are 0. In particular, $\textbf{\textit{A}}^n=\textbf{0}$. Exploiting this \textit{reduction property} of the matrix $\textbf{\textit{A}}$, we propose the following two solutions as ansatz for Eq.~(\ref{eq:04}),
\begin{subequations} \label{eq:xieta}
\begin{eqnarray}
~&& \text{\boldmath$\xi$}(s) =   \underbrace{\left[ \sum_{k=1}^{n}  \underbrace{\sum_{j=1}^{k}C_{k,j}s^{j-1}}_{\psi_k(s)} \textbf{\textit{A}}^{k-1} \right]}_{\textbf{\textit{M}}(s)} \cdot \textbf{\textit{v}} , \label{eq:xi} \\
~&& \text{\boldmath$\eta$}(s) = \nonumber \\
~&& \underbrace{\left[ \sum_{k=1}^{n} (1+s)^{k-1} \left( d_k \ln(1+s)+f_k \right) \textbf{\textit{A}}^{k-1} \right]}_{ \text{\boldmath$\Lambda$}(s)} \cdot \textbf{\textit{w}}. \label{eq:eta}
\end{eqnarray}
\end{subequations}

Here, $C_{k,j}$, $d_k$, and $f_k$ are mathematical constants, independent of any physical parameter, and are to be determined from the condition that both $\text{\boldmath$\xi$}(s)$ and $\text{\boldmath$\eta$}(s)$ have to satisfy Eq.~(\ref{eq:04}). Note that, as is the usual convention, we define $\textbf{\textit{A}}^0$ to be the identity matrix $\textbf{\textit{I}}$. Also $\textbf{\textit{v}}$ and $\textbf{\textit{w}}$ are some arbitrary $n \times 1$ constant vectors subject to the boundary conditions only. The general solution will be the sum of these two individual solutions,
\begin{equation}\label{eq:x}
\textbf{\textit{x}}(s) = \text{\boldmath$\xi$}(s) + \text{\boldmath$\eta$}(s) .
\end{equation}

We can consider the mathematical constants $C_{k,j}$ as the elements of a matrix $\textbf{\textit{C}}$ of size $n \times n$. Similarly, we can consider the mathematical constants $d_k$ and $f_k$ as the elements of the $n \times 1$ vectors $\textbf{\textit{d}}$ and $\textbf{\textit{f}}$, respectively. Consider the expressions $\psi_k (s)$ that are identified in Eq.~(\ref{eq:xi}).  Each $\psi_k (s)$ is a polynomial of order $k-1$ and can be considered as an element of a vector $ \text{\boldmath$\psi$}(s)$. This vector $\text{\boldmath$\psi$}(s)$ represents a special class of polynomials in variable $s$ and can be concisely described as,
\begin{equation}\label{eq:psi}
\text{\boldmath$\psi$}(s) = \textbf{\textit{C}} \cdot \text{\boldmath$\sigma$}(s).
\end{equation}

Here $\text{\boldmath$\sigma$}(s)$ is an $n \times 1$ vector whose $k$th element is defined as,
\begin{equation}\label{eq:sigma}
\sigma_k(s) \equiv s^{k-1}.
\end{equation}

Let us first determine the matrix  $\textbf{\textit{C}}$. Without any loss of generality, we can choose,
\begin{subequations} \label{eq:1C}
\begin{eqnarray}
C_{1,1}=~&&1, \\
C_{k,1}=~&&0, k>1,\\
C_{k,j}=~&&0, j>k.
\end{eqnarray}
\end{subequations}

Let us determine other $C_{k,j}$'s by substituting $\text{\boldmath$\xi$}(s)$ from Eq.~(\ref{eq:xi}) as $\textbf{\textit{x}}(s)$ in Eq.~(\ref{eq:04}) and comparing the coefficients of the expression $s^{j-2} \textbf{\textit{A}}^{k-1} \cdot \textbf{\textit{v}}$.
\begin{multline} \label{eq:Ckj}
\left[C_{k,j+1}s^j \textbf{\textit{A}}^{k-1} \cdot \textbf{\textit{v}} \right]'' + s \left[C_{k,j}s^{j-1} \textbf{\textit{A}}^{k-1} \cdot \textbf{\textit{v}} \right]'' \\
+\left[C_{k,j}s^{j-1} \textbf{\textit{A}}^{k-1} \cdot \textbf{\textit{v}} \right]'= \textbf{\textit{A}} \cdot \left[C_{k-1,j-1}s^{j-2} \textbf{\textit{A}}^{k-2} \cdot \textbf{\textit{v}} \right], \\
\Rightarrow C_{k,j} = \frac{1}{(j-1)^2} C_{k-1,j-1} - \frac{j}{j-1} C_{k,j+1}, j>1.
\end{multline}

Here are some general results from Eqs.~(\ref{eq:Ckj}),
\begin{subequations} \label{eq:Ckk}
\begin{eqnarray}
C_{k,k}=~&&\frac{1}{\left[ (k-1)! \right]^2},  \label{eq:Ckk} \\
C_{k,2}=~&&-2C_{k,3}~,~~k>2.  \label{eq:Ck2}
\end{eqnarray}
\end{subequations}

We can use Eq.~(\ref{eq:Ckk}) and then Eq.~(\ref{eq:Ckj}) recursively to find all non-trivial elements of the matrix $\textbf{\textit{C}}$ and hence the polynomial elements of the vector $\text{\boldmath$\psi$}(s)$. After the trivial $\psi_1 (s)=1$, some of the next few polynomials of this class, elements of $\text{\boldmath$\psi$}(s)$, are given as follows,
\begin{subequations}  \label{eq:psis}
\begin{eqnarray}
\psi_2=~&&s,\\
\psi_3=~&&(-2s+s^2)/4,\\
\psi_4=~&&(12s-6s^2+s^3)/36,\\
\psi_5=~&&(-132s+66s^2-12s^3+s^4)/576.
\end{eqnarray}
\end{subequations}

Similarly, let us determine the elements of the vectors $\textbf{\textit{d}}$ and $\textbf{\textit{f}}$. First, without any loss of generality, we take,
\begin{subequations} \label{eq:d1f1}
\begin{eqnarray}
d_1=~&&1,  \label{eq:d1} \\
f_1=~&&0.  \label{eq:f1}
\end{eqnarray}
\end{subequations}

Then we determine the other elements of $\textbf{\textit{d}}$ and $\textbf{\textit{f}}$ by substituting  $\text{\boldmath$\eta$}(s)$ from Eq.~(\ref{eq:eta}) as $\textbf{\textit{x}}(s)$ in Eq.~(\ref{eq:04}). After simplification, we get,
\begin{multline} \label{eq:dkfkcomp}
(1+s)^{k-2}\biggl( (k-1)^2 \bigl(d_k \ln(1+s) + f_k \bigr) +2(k-1)d_k \biggr) \\
= (1+s)^{k-2}\bigl(d_{k-1} \ln(1+s) + f_{k-1} \bigr).
\end{multline}

Comparing the coefficients of $(1+s)^{k-2} \ln(1+s)$ and $(1+s)^{k-2}$ in Eq.~(\ref{eq:dkfkcomp}) yields,
\begin{subequations} \label{eq:dkfk}
\begin{eqnarray}
d_k=~&&\frac{1}{\left[ (k-1)! \right]^2}=C_{k,k}\ ,  \label{eq:dk} \\
f_k=~&&\frac{1}{(k-1)^2} f_{k-1} - \frac{2}{k-1} d_k\ ,\ k>1.  \label{eq:fk}
\end{eqnarray}
\end{subequations}

It is important to note that since the elements of the matrix $\textbf{\textit{C}}$ and the vectors $\textbf{\textit{f}}$ and $\textbf{\textit{g}}$ are just mathematical constants, they can be determined a priori and stored in the memory for the later use for any instability problem. However, one has to carefully consider the possibility that reading an already calculated value from the memory could be more time consuming than calculating that value on the run.

Finally, let us determine the vectors $\textbf{\textit{v}}$ and $\textbf{\textit{w}}$ for a particular solution corresponding to the initial conditions. First, consider the explicit form of the general solution as a function of $s$, 
\begin{multline} \label{eq:xexplicit}
\textbf{\textit{x}}(s)=\sum_{k=1}^{n} \sum_{j=1}^{k}C_{k,j}s^{j-1} \textbf{\textit{A}}^{k-1} \cdot \textbf{\textit{v}} \\
+\sum_{k=1}^{n}(1+s)^{k-1} \bigl(d_k \ln(1+s) + f_k \bigr) \textbf{\textit{A}}^{k-1} \cdot \textbf{\textit{w}}.
\end{multline}

The initial conditions are given as,
\begin{subequations} \label{eq:qp}
\begin{eqnarray}
\textbf{\textit{q}} \equiv ~&&\textbf{\textit{x}}(0) =  \textbf{\textit{v}} + \underbrace{ \left[ \sum_{k=2}^{n} f_k \textbf{\textit{A}}^{k-1} \right] }_{\textbf{\textit{R}}} \cdot \textbf{\textit{w}} , \label{eq:q} \\
\textbf{\textit{p}} \equiv ~&&\textbf{\textit{x}}'(0)  =\frac{E(0)}{G} \left[ \frac{d\textbf{\textit{x}}(z)}{dz} \right]_{z=0}, \nonumber \\
~&& =\underbrace{\left[ \sum_{k=2}^{n} C_{k,2} \textbf{\textit{A}}^{k-1} \right]}_{\textbf{\textit{T}}} \cdot \textbf{\textit{v}} + \textbf{\textit{w}} \nonumber \\
~&& +\underbrace{\left[ \sum_{k=2}^{n} \bigl(d_k+(k-1)f_k \bigr) \textbf{\textit{A}}^{k-1} \right]}_{\textbf{\textit{U}}} \cdot \textbf{\textit{w}}. \label{eq:p}
\end{eqnarray}
\end{subequations}

Before proceeding further, let us define two more vectors that would serve as an algorithmic tool in the determination of the vectors $\textbf{\textit{v}}$ and $\textbf{\textit{w}}$.
\begin{subequations} \label{eq:ghdef}
\begin{eqnarray}
\textbf{\textit{g}}\equiv~&&\textbf{\textit{v}}-\textbf{\textit{q}},  \label{eq:gdef} \\
\textbf{\textit{h}}\equiv~&&\textbf{\textit{w}}-\textbf{\textit{p}}.  \label{eq:hdef}
\end{eqnarray}
\end{subequations}

Using Eq.~(\ref{eq:ghdef}), we can write the following equations for the vectors $\textbf{\textit{g}}$ and $\textbf{\textit{h}}$.
\begin{subequations}  \label{eq:gh0}
\begin{eqnarray}
\textbf{\textit{g}}=~&&-\textbf{\textit{R}}\cdot(\textbf{\textit{p}}+\textbf{\textit{h}}),  \label{eq:g0} \\
\textbf{\textit{h}}=~&&-\textbf{\textit{T}}\cdot(\textbf{\textit{q}}+\textbf{\textit{g}})-\textbf{\textit{U}}\cdot(\textbf{\textit{p}}+\textbf{\textit{h}}).  \label{eq:h}
\end{eqnarray}
\end{subequations}

Apparently, the vectors $\textbf{\textit{g}}$ and $\textbf{\textit{h}}$ in Eq.~(\ref{eq:gh0}) seem to be implicitly defined. Actually, however, these are very explicit relations. To see this, consider the matrices $\textbf{\textit{R}}$, $\textbf{\textit{T}}$, and $\textbf{\textit{U}}$, as identified in Eq.~(\ref{eq:qp}). Note that in the $k$th row of these matrices, all the elements from column number $k$ to $n$ are zero. Therefore,
\begin{equation}
g_1=h_1=0.
\end{equation}

Furthermore, for $k>1$, the $k$th elements of both the vectors $\textbf{\textit{g}}$ and $\textbf{\textit{h}}$ only depend on their first $k-1$ elements as follows,
\begin{subequations} \label{eq:gh}
\begin{eqnarray} 
g_k=~&&-\textbf{\textit{R}}_{k,1:k-1}\cdot(\textbf{\textit{p}}+\textbf{\textit{h}})_{1:k-1},  \label{eq:g} \\
h_k=~&&-\textbf{\textit{T}}_{k,1:k-1}\cdot(\textbf{\textit{q}}+\textbf{\textit{g}})_{1:k-1} \nonumber \\
~&&-\textbf{\textit{U}}_{k,1:k-1}\cdot(\textbf{\textit{p}}+\textbf{\textit{h}})_{1:k-1}.  \label{eq:h}
\end{eqnarray}
\end{subequations}

Here, the notation in the subscript $j:k$ means element number $j$ to $k$ for the particular row or column being considered. Before proceeding, let us write the explicit equations of various matrices that are used in the solution of Eq.~(\ref{eq:04}).
\begin{subequations} \label{eq:matrices}
\begin{eqnarray}
~&&\textbf{\textit{M}}(s)=\textbf{\textit{I}}+\sum_{k=2}^{n}\psi_k(s)\textbf{\textit{A}}^{k-1},  \label{eq:M} \\
~&&\text{\boldmath$\Lambda$}(s)=\ln(1+s)\textbf{\textit{I}} \nonumber \\
~&&+\sum_{k=2}^{n}(1+s)^{k-1}\bigl( C_{k,k} \ln(1+s)+f_k \bigr) \textbf{\textit{A}}^{k-1},  \label{eq:Lambda} \\
~&&\textbf{\textit{R}}=\sum_{k=2}^{n}f_k \textbf{\textit{A}}^{k-1},  \label{eq:R} \\
~&&\textbf{\textit{T}}=\sum_{k=2}^{n}C_{k,2} \textbf{\textit{A}}^{k-1},  \label{eq:T} \\
~&&\textbf{\textit{U}}=\sum_{k=2}^{n}\bigl( C_{k,k}+(k-1)f_k \bigr) \textbf{\textit{A}}^{k-1}.  \label{eq:U}
\end{eqnarray}
\end{subequations}

The final solution and its first derivative are given as,
\begin{subequations} \label{eq:xandxp}
\begin{eqnarray}
\textbf{\textit{x}}(s)~&&=\textbf{\textit{M}}(s) \cdot (\textbf{\textit{q}}+\textbf{\textit{g}}) + \text{\boldmath$\Lambda$}(s) \cdot (\textbf{\textit{p}}+\textbf{\textit{h}}), \label{eq:xfinal} \\
\textbf{\textit{x}}'(s)~&&=\textbf{\textit{M}}'(s) \cdot (\textbf{\textit{q}}+\textbf{\textit{g}}) + \text{\boldmath$\Lambda$}'(s) \cdot (\textbf{\textit{p}}+\textbf{\textit{h}}). \label{eq:xpfinal}
\end{eqnarray}
\end{subequations}

Here is a fast algorithm to obtain the solution of Eq.~(\ref{eq:04}) and this also summarizes our discussion so far.

\begin{tabbing}
\line(1,0){240}\\
\textbf{Algorithm: General Case}\\
Input \= the number of bunches $n$, the initial\\
\> energy (including the rest mass energy) $E(0)$,\\
\> the constant energy gradient $G$, and\\
\> the final longitudinal position $z$;\\
Input the $n \times 1$ vectors of bunch charges and relative\\
\> longitudinal positions: $\textbf{\textit{c}}$, $\textbf{\textit{z}}$;\\
Input the $n \times 1$ vectors defining the initial conditions: \\
\> $\textbf{\textit{q}}=\textbf{\textit{x}}(0)$, $\textbf{\textit{p}}=\frac{E(0)}{G} \left[ \frac{d\textbf{\textit{x}}(z)}{dz} \right]_{z=0}$;\\
\\
Calculate the normalized quantity: $s=\frac{G\ z}{E(0)}$;\\
Define an $n \times 1$ normalized vector \text{\boldmath$\sigma$} with elements:\\
\> $\sigma_k=s^{k-1}$;\\
Initialize an $n \times n$ normalized matrix: $\textbf{\textit{A}} = \textbf{0}$;\\
For \= $k$ from 2 to $n$,\\
\> For \= $j$ from 1 to $k-1$,\\
\> \> $A_{k,j}=\frac{e E(0)}{G^2}c_j W(z_{j}-z_{k})$;\\
\> End\\
End\\
 \\
Initialize $1 \times n$ vectors: $\textbf{\textit{cold}}=\textbf{\textit{cnew}}=\textbf{0}$;\\
$cold_1=1$;\\
Initialize: $f=0$;\\
Initialize: $\psi=1$,  $\psi'=0$;\\
Initialize $n \times n$ matrices: $\textbf{\textit{Akminus1}}=\textbf{\textit{M}}=\textbf{\textit{I}}$;\\
Initialize an $n \times n$ matrix: $\textbf{\textit{M}}'=\textbf{0}$;\\
Initialize an $n \times n$ matrix: $\text{\boldmath$\Lambda$}=\ln(1+s)\textbf{\textit{I}}$;\\
Initialize an $n \times n$ matrix: $\text{\boldmath$\Lambda$}'=\frac{1}{1+s}\textbf{\textit{I}}$;\\
Initialize $n \times n$ matrices: $\textbf{\textit{R}}=\textbf{\textit{T}}=\textbf{\textit{U}}=\textbf{0}$;\\
Initialize $n \times 1$ vectors: $\textbf{\textit{g}}=\textbf{\textit{h}}=\textbf{0}$;\\
\\
For \= $k$ from 2 to $n$,\\
\> $cnew_k = \frac{1}{(k-1)^2}cold_{k-1}$;\\
\> For \= $j$ decreasing from $k-1$ to $2$,\\
\> \> $cnew_j=\frac{1}{(j-1)^2}cold_{j-1}-\frac{j}{j-1}cnew_{j+1}$;\\
\> End\\
\> $f=\frac{1}{(k-1)^2}f-\frac{2}{k-1}cnew_k$;\\
\> $\psi = \textbf{\textit{cnew}}_{2:k} \cdot \text{\boldmath$\sigma$}_{2:k}$;\\
\> $\psi' = \textbf{\textit{D}}(1:k-1) \cdot \textbf{\textit{cnew}}_{2:k} \cdot \text{\boldmath$\sigma$}_{1:k-1}$;\\
\> \> Comment: $\textbf{\textit{D}}(1:k-1)$ is a diagonal matrix\\
\> \> with diagonal entries $1,2,...,k-1$.\\
\> $\textbf{\textit{Akminus1}} = \textbf{\textit{A}} \cdot \textbf{\textit{Akminus1}}$;\\
\> $\textbf{\textit{M}} +=  \psi \textbf{\textit{Akminus1}}$;\\
\> $\textbf{\textit{M}}' +=  \psi' \textbf{\textit{Akminus1}}$;\\
\> $\text{\boldmath$\Lambda$} += $\\
\> $\quad (1+s)^{k-1} \big( cnew_k \ln(1+s)+f\big)\textbf{\textit{Akminus1}}$;\\
\> $\text{\boldmath$\Lambda$}'+= (1+s)^{k-2}$\\
\> $\quad \big( (k-1)(cnew_k \ln(1+s)+f)+cnew_k\big)$\\
\> $\quad \textbf{\textit{Akminus1}}$;\\
\> $\textbf{\textit{R}} +=  f \textbf{\textit{Akminus1}}$;\\
\> $\textbf{\textit{T}} +=  cnew_2 \textbf{\textit{Akminus1}}$;\\
\> $\textbf{\textit{U}} +=  \big( cnew_k+(k-1)f \big) \textbf{\textit{Akminus1}}$;\\
\> $g_k = -\textbf{\textit{R}}_{k,1:k-1} \cdot (\textbf{\textit{p}}+\textbf{\textit{h}})_{1:k-1}$;\\
\> $h_k = -\textbf{\textit{T}}_{k,1:k-1} \cdot (\textbf{\textit{q}}+\textbf{\textit{g}})_{1:k-1}$\\
\> $\quad \quad \ -\textbf{\textit{U}}_{k,1:k-1} \cdot (\textbf{\textit{p}}+\textbf{\textit{h}})_{1:k-1}$;\\
\> $cold_{1:k}=cnew_{1:k}$;\\
End\\
\\
Calculate and output:\\
\> $\textbf{\textit{x}}=\textbf{\textit{M}} \cdot (\textbf{\textit{q}}+\textbf{\textit{g}}) + \text{\boldmath$\Lambda$}\cdot (\textbf{\textit{p}}+\textbf{\textit{h}})$,\\
\> $\textbf{\textit{x}}'=\textbf{\textit{M}}' \cdot (\textbf{\textit{q}}+\textbf{\textit{g}}) + \text{\boldmath$\Lambda$}'\cdot (\textbf{\textit{p}}+\textbf{\textit{h}})$;\\
\line(1,0){240}
\end{tabbing}

\section{\label{sec:specific} Specific and Common Case}
Consider the specific and more common case when all the bunches have the same charge, $c_k=c$ , and the longitudinal separation between any two consecutive bunches is the same, $z_k-z_{k-1}=\Delta z$. In this case, there are only $n$ fundamental entries in the matrix \textbf{\textit{A}} given as follows,
\begin{subequations} \label{eq:Aanda}
\begin{eqnarray}
A_{k,j} = ~&&\left\{ \begin{array}{ll}
            0, &j > k-1, \\
            a_{k-j}, &\text{otherwise,}
        \end{array}
 \right.  \label{eq:Aspecial} \\
a_k \equiv ~&& \frac{e\ E(0)}{G^2} c\ W(k\ \Delta z). \label{eq:afund}
\end{eqnarray}
\end{subequations}

Following the definitions provided in the Appendix, we note that $\textbf{\textit{A}}$, in this specific case, is an omega matrix of degree $1$. Therefore, any power matrix $\textbf{\textit{A}}^r$ is an omega matrix of degree $r$. Similarly, the matrices $\textbf{\textit{M}}$ and $\text{\boldmath$\Lambda$}$ are omega matrices of degree $0$, while the matrices $\textbf{\textit{R}}$, $\textbf{\textit{T}}$, and $\textbf{\textit{U}}$ are omega matrices of degree $1$ . Thus, in this specific case, we can describe all these matrices very concisely by their corresponding fundamental vectors. Any matrix $\textbf{\textit{A}}^{r}$ is fully described by its fundamental vector $\textbf{\textit{a}}^{(r)}$ of size $(n- r) \times 1$. Consider the $(n-1) \times 1$ vector $\textbf{\textit{a}}$ whose elements are defined by Eq.~(\ref{eq:afund}). According to Eq.~(\ref{eq:Aspecial}), $\textbf{\textit{a}}$ is the fundamental vector of the matrix $\textbf{\textit{A}}$. Using the definition of the convolution function as defined in the Appendix, the fundamental vector of $\textbf{\textit{A}}^{r+1}$ can be obtained in terms of the fundamental vectors of $\textbf{\textit{A}}$ and $\textbf{\textit{A}}^r$ as follows,
\begin{equation} \label{eq:arplus1}
\textbf{\textit{a}}^{(r+1)}=\mathrm{Conv}\left[\textbf{\textit{a}},\textbf{\textit{a}}^{(r)},n-r-1\right] .
\end{equation}

Note that $\textbf{\textit{a}}^{(0)}=i^{(0)}$ and that  $\textbf{\textit{a}}^{(1)}= \textbf{\textit{a}}$. Let us now write the fundamental vectors of our matrices of interest. We will use the corresponding lower case letter for the fundamental vectors.
\begin{subequations} \label{eq:fundvect}
\begin{eqnarray}
~&&\textbf{\textit{m}}(s)=\textbf{\textit{i}}^{(0)}+\sum_{k=2}^{n}\psi_k(s)\textbf{\textit{a}}^{(k-1)},  \label{eq:mfund} \\
~&&\text{\boldmath$\lambda$}(s)=\ln(1+s)\textbf{\textit{i}}^{(0)} \nonumber \\
~&&+\sum_{k=2}^{n}(1+s)^{k-1}\bigl( C_{k,k} \ln(1+s)+f_k \bigr) \textbf{\textit{a}}^{(k-1)},  \label{eq:lambdafund} \\
~&&\textbf{\textit{r}}=\sum_{k=2}^{n}f_k \textbf{\textit{a}}^{(k-1)},  \label{eq:rfund} \\
~&&\textbf{\textit{t}}=\sum_{k=2}^{n}C_{k,2} \textbf{\textit{a}}^{(k-1)},  \label{eq:tfund} \\
~&&\textbf{\textit{u}}=\sum_{k=2}^{n}\bigl( C_{k,k}+(k-1)f_k \bigr) \textbf{\textit{a}}^{(k-1)}.  \label{eq:ufund}
\end{eqnarray}
\end{subequations} 

Consider the following,
\begin{equation} \label{eq:Rk}
\textbf{\textit{R}}_{k,1:k-1}=\textbf{\textit{R}}_{n,n-k+1:n-1}=\textbf{\textit{r}}_{k-1:1} \equiv \mathrm{Reverse}[\textbf{\textit{r}}_{1:k-1}] .
\end{equation}

Like Eq.~(\ref{eq:Rk}), we can use similar simplifications for $\textbf{\textit{T}}_{k,1:k-1}$ and $\textbf{\textit{U}}_{k,1:k-1}$ in Eq.~(\ref{eq:gh}) which can be written as follows in terms of the fundamental vectors.
\begin{subequations} \label{eq:ghspecial}
\begin{eqnarray}
g_k=~&&-\mathrm{Reverse}[\textbf{\textit{r}}_{1:k-1}]\cdot(\textbf{\textit{p}}+\textbf{\textit{h}})_{1:k-1},  \label{eq:gspecial} \\
h_k=~&&-\mathrm{Reverse}[\textbf{\textit{t}}_{1:k-1}]\cdot(\textbf{\textit{q}}+\textbf{\textit{g}})_{1:k-1} \nonumber \\
~&&-\mathrm{Reverse}[\textbf{\textit{u}}_{1:k-1}]\cdot(\textbf{\textit{p}}+\textbf{\textit{h}})_{1:k-1}.  \label{eq:hspecial}
\end{eqnarray}
\end{subequations}

Finally, utilizing the result of Eq.~(\ref{eq:omega0beta}) we can write the final solution and its derivative as,
\begin{subequations} \label{eq:xandxpspecial}
\begin{eqnarray}
\textbf{\textit{x}}(s)~&&=\mathrm{Conv}[\textbf{\textit{m}}(s),(\textbf{\textit{q}}+\textbf{\textit{g}}),n] \nonumber \\
~&&+ \mathrm{Conv}[\text{\boldmath$\lambda$}(s),(\textbf{\textit{p}}+\textbf{\textit{h}}),n], \label{eq:xfinalspecial} \\
\textbf{\textit{x}}'(s)~&&=\mathrm{Conv}[\textbf{\textit{m}}'(s),(\textbf{\textit{q}}+\textbf{\textit{g}}),n] \nonumber \\
~&&+ \mathrm{Conv}[\text{\boldmath$\lambda$}'(s),(\textbf{\textit{p}}+\textbf{\textit{h}}),n]. \label{eq:xpfinalspecial}
\end{eqnarray}
\end{subequations}

Since in this specific case we can describe all matrices by their fundamental vectors, we can utilize this fact to have a much faster algorithm with far less memory-reads and arithmetic operations. The following is a fast algorithm that employs the fundamental vectors of the matrices to solve Eq.~(\ref{eq:04}) for this specific case.

\begin{tabbing}
\line(1,0){240}\\
\textbf{Algorithm: Specific and Common Case}\\
Input \= the number the number of bunches $n$, the initial\\
\> energy (including the rest mass energy) $E(0)$,\\
\> the constant energy gradient $G$,\\
\> the final longitudinal position $z$,\\
\> the bunch charge $c$, and\\
\> the interbunch longitudinal separation $\Delta z$;\\
Input the $n \times 1$ vectors defining the initial conditions: \\
\> $\textbf{\textit{q}}=\textbf{\textit{x}}(0)$, $\textbf{\textit{p}}=\frac{E(0)}{G} \left[ \frac{d\textbf{\textit{x}}(z)}{dz} \right]_{z=0}$;\\
\\
Calculate the normalized quantity: $s=\frac{G\ z}{E(0)}$;\\
Define an $n \times 1$ normalized vector \text{\boldmath$\sigma$} with elements:\\
\> $\sigma_k=s^{k-1}$;\\
Initialize an $(n-1) \times 1$ normalized vector: $\textbf{\textit{a}} = \textbf{0}$;\\
For \= $k$ from 1 to $n-1$,\\
\> $a_k = \frac{e\ E(0)}{G^2} c\ W(k\ \Delta z)$;\\
End\\
 \\
Initialize $1 \times n$ vectors: $\textbf{\textit{cold}}=\textbf{\textit{cnew}}=\textbf{0}$;\\
$cold_1=1$;\\
Initialize: $f=0$;\\
Initialize: $\psi=1$,  $\psi'=0$;\\
Initialize $n \times 1$ vectors: $\textbf{\textit{akminus1}}=\textbf{\textit{m}}=\text{\boldmath$\lambda$}=\textbf{0}$;\\
$akminus1_1=m_1=1$;\\
$\lambda_1=\ln(1+s)$;\\
Initialize $n \times 1$ vectors: $\textbf{\textit{m}}'=\text{\boldmath$\lambda$}'=\textbf{0}$;\\
$\lambda'_1=\frac{1}{1+s}$;\\
Initialize $(n-1) \times 1$ vectors: $\textbf{\textit{r}}=\textbf{\textit{t}}=\textbf{\textit{u}}=\textbf{0}$;\\
Initialize $n \times 1$ vectors: $\textbf{\textit{g}}=\textbf{\textit{h}}=\textbf{0}$;\\
\\
For \= $k$ from 2 to $n$,\\
\> $cnew_k = \frac{1}{(k-1)^2}cold_{k-1}$;\\
\> For \= $j$ decreasing from $k-1$ to $2$,\\
\> \> $cnew_j=\frac{1}{(j-1)^2}cold_{j-1}-\frac{j}{j-1}cnew_{j+1}$;\\
\> End\\
\> $f=\frac{1}{(k-1)^2}f-\frac{2}{k-1}cnew_k$;\\
\> $\psi = \textbf{\textit{cnew}}_{2:k} \cdot \text{\boldmath$\sigma$}_{2:k}$;\\
\> $\psi' = \textbf{\textit{D}}(1:k-1) \cdot \textbf{\textit{cnew}}_{2:k} \cdot \text{\boldmath$\sigma$}_{1:k-1}$;\\
\> \> Comment: $\textbf{\textit{D}}(1:k-1)$ is a diagonal matrix\\
\> \> with diagonal entries $1,2,...,k-1$.\\
\> $\textbf{\textit{akminus1}} = \mathrm{Conv}\left[\textbf{\textit{a}},\textbf{\textit{akminus1}},n-k+1\right] $;\\
\> $\textbf{\textit{m}}_{k:n} +=  \psi \textbf{\textit{akminus1}}$;\\
\> $\textbf{\textit{m}}'_{k:n} +=  \psi' \textbf{\textit{akminus1}}$;\\
\> $\text{\boldmath$\lambda$}_{k:n} += (1+s)^{k-1} \big( cnew_k \ln(1+s)+f\big)$\\
\> $\quad \textbf{\textit{akminus1}}$;\\
\> $\text{\boldmath$\lambda$}'_{k:n}+= (1+s)^{k-2}$\\
\> $\quad \big( (k-1)(cnew_k \ln(1+s)+f)+cnew_k\big)$\\
\> $\quad \textbf{\textit{akminus1}}$;\\
\> $\textbf{\textit{r}}_{k-1:n-1} +=  f \textbf{\textit{akminus1}}$;\\
\> $\textbf{\textit{t}}_{k-1:n-1} +=  cnew_2 \textbf{\textit{akminus1}}$;\\
\> $\textbf{\textit{u}}_{k-1:n-1} +=  \big( cnew_k+(k-1)f \big) \textbf{\textit{akminus1}}$;\\
\> $g_k = -\mathrm{Reverse}[\textbf{\textit{r}}_{1:k-1}]\cdot(\textbf{\textit{p}}+\textbf{\textit{h}})_{1:k-1}$;\\
\> $h_k = -\mathrm{Reverse}[\textbf{\textit{t}}_{1:k-1}]\cdot(\textbf{\textit{q}}+\textbf{\textit{g}})_{1:k-1}$\\
\> $\quad \quad \ -\mathrm{Reverse}[\textbf{\textit{u}}_{1:k-1}]\cdot(\textbf{\textit{p}}+\textbf{\textit{h}})_{1:k-1}$;\\
\> $cold_{1:k}=cnew_{1:k}$;\\
End\\
\\
Calculate and output:\\
\> $\textbf{\textit{x}}= \mathrm{Conv}\left[\textbf{\textit{m}},(\textbf{\textit{q}}+\textbf{\textit{g}}),n\right] +  \mathrm{Conv}\left[ \text{\boldmath$\lambda$},(\textbf{\textit{p}}+\textbf{\textit{h}}),n\right]$,\\
\> $\textbf{\textit{x}}'= \mathrm{Conv}\left[\textbf{\textit{m}}',(\textbf{\textit{q}}+\textbf{\textit{g}}),n\right] +  \mathrm{Conv}\left[ \text{\boldmath$\lambda$}',(\textbf{\textit{p}}+\textbf{\textit{h}}),n\right]$;\\
\line(1,0){240}
\end{tabbing}

\section{\label{sec:example} Example}

Consider an X-band electron accelerator with three dominant dipole modes. The resonant frequency $F\!R$, quality factor $Q\!F$, and the transverse loss factor $K\!F$ of these modes are given in Table~\ref{tab:01}. The last column lists the the form factor $F\!F$ corresponding to these modes for a gaussian bunch size $d=0.6$~mm. The form factor is given as $F\!F=\exp \left(-\frac{1}{2}(\frac{2 \pi d\ F\!R}{c_0})^2 \right)$ where $c_0$ is the speed of light.

\begin{table}[ht]
\caption{Dominant Dipole Modes} 
\centering 
\begin{tabular}{c c c c c} 
\hline\hline 
Mode & $F\!R$ [GHz] & $Q\!F$ & $K\!F$ [MV/pC/m/m]  & $F\!F$ \\ [0.5ex] 
\hline 
1 & 16.5 & 9,857 & 0.004  & 0.979\\ 
2 & 19.4 & 12,566 & 0.086 & 0.971\\
3 & 25.9 & 10,790 & 0.139 & 0.948\\ [1ex] 
\hline 
\end{tabular}
\label{tab:01} 
\end{table}

\begin{figure}[h!]
\includegraphics{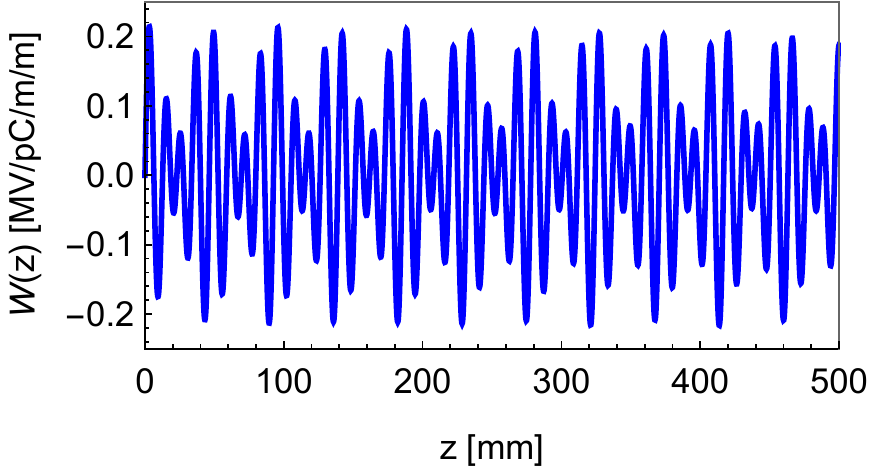}
\caption{\label{fig:wake} The wakefield function given in Eq.~\ref{eq:W}.}
\end{figure}

The transverse wakefield function is given in Eq.~(\ref{eq:W}) and is plotted in Fig.~\ref{fig:wake}
\begin{equation}\label{eq:W}
W(z)=\sum_{i=1}^{3} F\!F_i\ K\!F_i \sin\left(\frac{2\pi\ F\!R_i}{c_0}z\right) \exp \left( -\frac{\pi\ F\!R_i}{c_0\ Q\!F_i}z \right).
\end{equation}

The frequency of this accelerator is 11.424~GHz. All rf buckets are filled, therefore, the interbunch separation is 26.24~mm.  Assuming a 1~$\mu$s pulse width, the total number of bunches in the pulse width are: $n=11,424$. The bunches are linearly accelerated from an initial total energy $E(0)=(0.5+0.511)$~MeV through a constant gradient of $G=20$~MV/m.  All the bunches are assumed to have the same charge. We consider two cases of beam current: 100~mA and 200~mA. Thus, the bunch charge for these two cases is 8.75~pC and 17.5~pC, respectively.

\begin{figure}[t]
\includegraphics{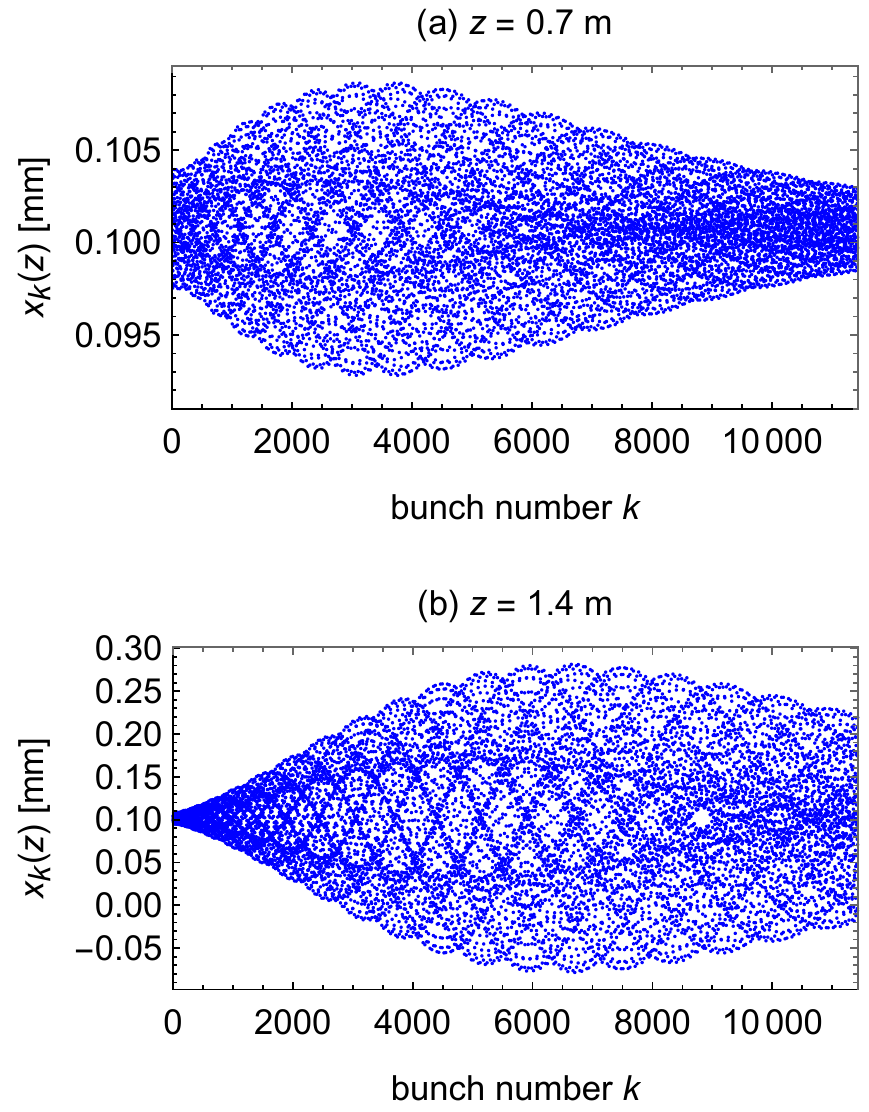}
\caption{\label{fig:x100} The transverse positions of the particles at two different longitudinal positions for the example in Sec. \ref{sec:example} assuming a beam current of 100~mA.}
\end{figure}

Let us assume that the initial transverse positions and deflections of the particles are given as $x_k(0)=0.1$~mm and $\left[ \frac{dx_k(z)}{dz} \right]_{z=0}=0$, respectively. Since this is the case of uniform bunch charge and uniform interbunch separation, we can use the specific case algorithm to calculate the transverse positions of the particles at various longitudinal positions. Figure~\ref{fig:x100} shows the transverse positions of the particles at two different longitudinal positions, $z=0.7$~m and $z=1.4$~m, for the case of 100~mA beam current. We see that in this case the bunches do not deviate more than 0.3~mm from the axis even after traveling a distance of 1.4~m through the accelerator.

\begin{figure}[t]
\includegraphics{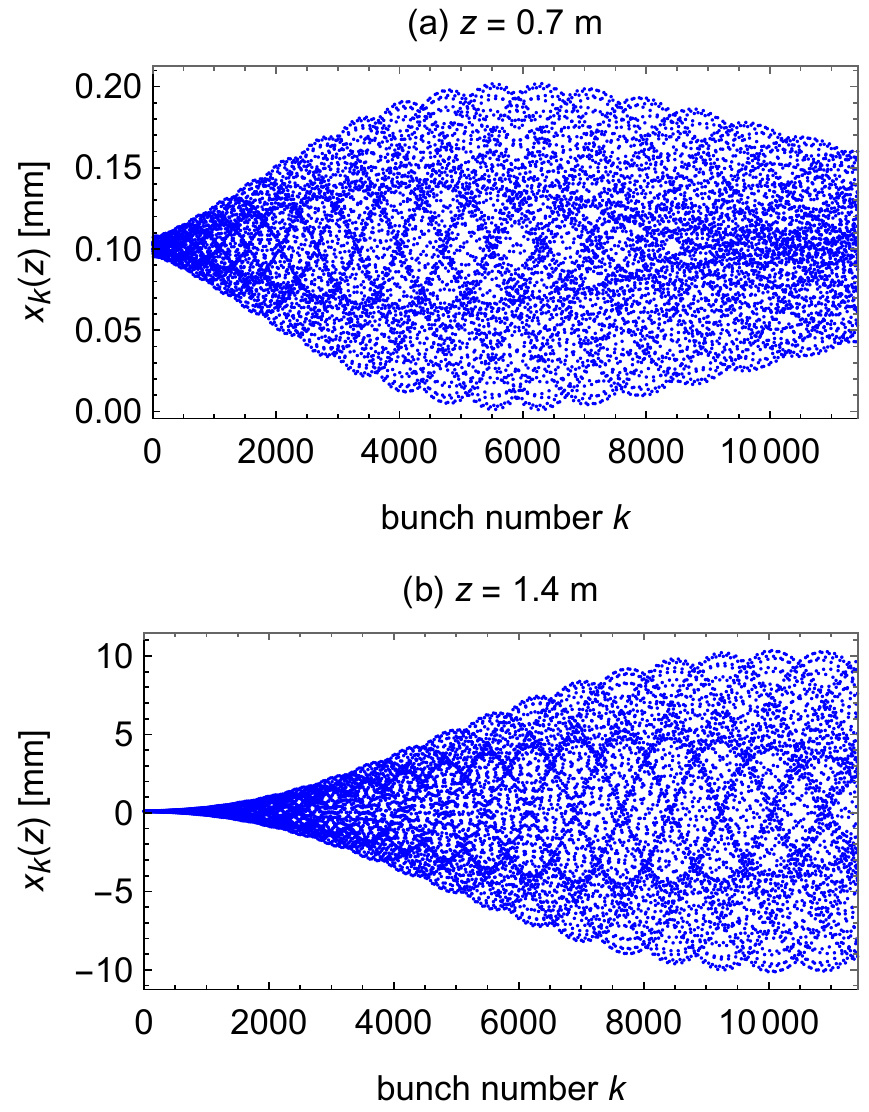}
\caption{\label{fig:x200} The transverse positions of the particles at two different longitudinal positions for the example in Sec. \ref{sec:example} assuming a beam current of 200~mA.}
\end{figure}

Figure~\ref{fig:x200} shows similar results for the case of 200~mA beam current. Here, however, we observe a huge increase in the transverse coordinates of the bunches at $z=1.4$~m. It turns out that, if the aperture radius of the accelerator is 1~mm, then about two thirds of the bunches in this 200~mA beam are lost somewhere between $z=0.7$~m and $z=1.4$~m.

The results shown in Fig.~\ref{fig:x100} and Fig. \ref{fig:x200} were obtained with an implementation of the specific case algorithm in Mathematica\textsuperscript{\textregistered} 12 on a typical laptop  (Intel\textsuperscript{\textregistered}  Core\textsuperscript{TM} i7-9805H CPU, 2.60~GHz,  64~GB). The code was compiled using the option $CompilationTarget \rightarrow ''\!C''$.

A comment is in order regarding the \textit{for loop} in our algorithms in which $k$ runs from 2 to $n$. Here, we update the matrices $\textbf{\textit{M}}(s)$, $\textbf{\textit{M}}'(s)$, $\text{\boldmath$\Lambda$}(s)$, $\text{\boldmath$\Lambda$}'(s)$, $\textbf{\textit{R}}$, $\textbf{\textit{T}}$, and $\textbf{\textit{U}}$, or their corresponding fundamental vectors in the specific case, with each iteration. Each iteration corresponds to adding a term proportional to $\textbf{\textit{A}}^{k-1}$, or $\textbf{\textit{a}}^{(k-1)}$ in the specific case. If the trajectories are stable, as is the case in the example mentioned here, it turns out that these matrices converge very quickly and thus, for all practicle puposes, we need not to update these matrices or their corresponding fundamental vectors beyond a certain value of the loop variable $k$. This helps make the implementaiton of our algorithms even faster. In this example, when we updated the matrices $n=11,424$ times, as per the rigorous solution, the code took about 14 minutes to calculate the resutls for each plot. However, when we updated the fundamental vectors only 100 times, the results were obtained within just about a second of CPU time while the maximum relative error was only of the order $<10^{-12}$. In general, when the incremental updates reduce to the same order as that of the machine accuracy, there is no practicle need to update the matrices or their fundamental vectors with such accuracy-limited and time-consuming increments.

\section{\label{sec:conc} Conclusion }
We have presented an exact solution for the evolution of the transverse coordinates of ultrarelativistic point-like bunches of charge particles that are being accelerated through a constant energy gradient while being under the influence of the transverse wakefields of the leading bunches. We have shown that for the specific but more common case of uniform bunch charge and uniform interbunch longitudinal separation, the matrices of the general solution can be represented by vectors and hence the time-consuming matrix multiplications are replaced by fast vector convolutions. We have summarized the algorithms for both the general and the specific case. We have also provided an example demonstrating the usage of our fast algorithms for the practicle cases of exploring multibunch instabilities in particle accelerators.

This method could also be used for the analysis of the multibunch effects in an accelerator beam line with a non-uniform gradient. For such a problem, the beam line will be considered as comprising of various sections of constant energy gradient. This method will then be applied to each section in a cascaded manner.

This approach is also applicable for the analysis of the single-bunch beam break up (SBBU). In the case of SBBU, we  assume that the  bunch is comprised of several transverse slices. Then we use the short-range transverse wakefield of a single slice to simulate the evolution of the bunch shape. The practical simulations of single-bunch dynamics would require the bunch to be divided into hundreds of slices. Such a simulation will produce more precise results than the simpler two particle model \cite{Balakin:1978, Wangler:2008a, Wiedemann:2007}. Besides being more precise, this multiple-slice approach will still be fast enough (a few seconds on a typical computer). 

The rigorous analytical solution presented in this paper is not only free from numerical approximations but is also a much faster tool when compared with the numerical differential equation solvers. Our specific and common case algorithm can analyze tens of thousands of bunches for the multibunch instability problem within only a a couple of seconds on a typical laptop. The Mathematica\textsuperscript{\textregistered} implementation of both algorithms, general and specific, can be obtained from the author.

\begin{acknowledgments}
The author is thankful to Emilio A. Nanni for providing the motivation that was inevitably needed to complete this manuscript. This work was supported under USA Department of Energy (DOE) Contract No. DE-AC02-76SF00515.
\end{acknowledgments}

\appendix*
\section{The Omega Matrices and Their Fundamental Vectors \label{sec:A}}
\subsection{Definitions \label{sec:A1}}
Our definition of the term omega matrix here will be different than another use of this term in the field of economics \cite{Langsen:1989}.
We define an \textit{omega matrix} $\text{\boldmath$\Omega$}^{(r)}$ of size $n$ and degree $r$, where $0 \leq r \leq n$, as an $n \times n$ square matrix whose elements are defined as follows,
\begin{equation} \label{eq:omegar}
\Omega^{(r)}_{k,j} = \left\{ \begin{array}{ll}
            0, &j > k-r, \\
            \omega^{(r)}_{k-j-r+1}, &\text{otherwise.}
        \end{array}
 \right. 
\end{equation}

We call $\omega^{(r)}_k$, where $1 \leq k \leq n-r$, as the \textit{fundamental elements} of the omega matrix. It should be noted that we are using an integer inside the paranthesis in the superscript to denote the degree of the omega matrix and this is to contrast it with the common meaning of the exponent notation. According to the definition given above, the omega matrix of size and degree $n$ is a trivial matrix $\text{\boldmath$\Omega$}^{(n)}=\textbf{0}$ . To illustrate the above definition, let us explicitly depict all non-trivial omega matrices of size $n=4$ and degree $r=0,1,2,3$. 
\begin{subequations} \label{eq:omegaex}
\begin{eqnarray}
\text{\boldmath$\Omega$}^{(0)} = ~&&\begin{pmatrix}
\omega^{(0)}_1 & 0 & 0 & 0\\
\omega^{(0)}_2 & \omega^{(0)}_1 & 0 & 0\\
\omega^{(0)}_3 & \omega^{(0)}_2 & \omega^{(0)}_1 & 0\\
\omega^{(0)}_4 & \omega^{(0)}_3 & \omega^{(0)}_2 & \omega^{(0)}_1
\end{pmatrix},  \label{eq:omega0} \\
\text{\boldmath$\Omega$}^{(1)} = ~&&\begin{pmatrix}
0 & 0 & 0 & 0\\
\omega^{(1)}_1 & 0 & 0 & 0\\
\omega^{(1)}_2 & \omega^{(1)}_1 & 0 & 0\\
\omega^{(1)}_3 & \omega^{(1)}_2 & \omega^{(1)}_1 & 0
\end{pmatrix},  \label{eq:omega1} \\
\text{\boldmath$\Omega$}^{(2)} = ~&&\begin{pmatrix}
0 & 0 & 0 & 0\\
0 & 0 & 0 & 0\\
\omega^{(2)}_1 & 0 & 0 & 0\\
\omega^{(2)}_2 & \omega^{(2)}_1 & 0 & 0
\end{pmatrix},  \label{eq:omega2} \\
\text{\boldmath$\Omega$}^{(3)} = ~&&\begin{pmatrix}
0 & 0 & 0 & 0\\
0 & 0 & 0 & 0\\
0 & 0 & 0 & 0\\
\omega^{(3)}_1 & 0 & 0 & 0
\end{pmatrix}.  \label{eq:omega3}
\end{eqnarray}
\end{subequations}

Note that the all the fundamental elements of the omega matrix of size $n$ and degree $r$ are present in the last row, from column number $1$ to $n-r$, other elements of this row being zero. All other rows can then be obtained by noticing that a certain $(k-1)$th row is the left-shifted version of the $k$th row. Thus, we can economically denote any omega matrix of size $n$ and degree $r$ by an $(n- r) \times 1$ vector $\text{\boldmath$\omega$}^{(r)}$ that contains its fundamental elements such that,
\begin{equation} \label{eq:fundvector}
\omega^{(r)}_k=\Omega^{(r)}_{n,n-r-k+1}.
\end{equation}

We call $\text{\boldmath$\omega$}^{(r)}$ as the \textit{fundamental vector} of the omega matrix $\text{\boldmath$\Omega$}^{(r)}$. Here are the fundamental vectors, respectively listed, corresponding to the omega matrices given in Eq.~(\ref{eq:omegaex}):
\begin{subequations} \label{eq:fundvectorex}
\begin{eqnarray}
\text{\boldmath$\omega$}^{(0)} = ~&&\begin{pmatrix}
\omega^{(0)}_1 & \omega^{(0)}_2 & \omega^{(0)}_3 & \omega^{(0)}_4
\end{pmatrix}^{T},  \\
\text{\boldmath$\omega$}^{(1)} = ~&&\begin{pmatrix}
\omega^{(1)}_1 & \omega^{(1)}_2 & \omega^{(1)}_3
\end{pmatrix}^{T},   \\
\text{\boldmath$\omega$}^{(2)} = ~&&\begin{pmatrix}
\omega^{(2)}_1 & \omega^{(2)}_2
\end{pmatrix}^{T},  \\
\text{\boldmath$\omega$}^{(3)} = ~&&\begin{pmatrix}
\omega^{(3)}_1 
\end{pmatrix}. 
\end{eqnarray}
\end{subequations}

The superscript $^T$ in Eq.~(\ref{eq:fundvectorex}) indicates the transpose operation. Note that the identity matrix $\textbf{\textit{I}}$ is an omega matrix of degree 0 and has a corresponding $n \times 1$ fundamental vector $\textbf{\textit{i}}^{(0)}$ defined as follows,
\begin{equation} \label{eq:i0}
i^{(0)}_k = \left\{ \begin{array}{ll}
            1, &k=1, \\
            0, &\text{otherwise.}
        \end{array}
 \right. 
\end{equation}

Another useful observation is that the sum of two omega matrices, one of degree $q$ and the other of degree $r \geq q$, is also an omega matrix and its degree is $r$.

\subsection{Product of two omega matrices \label{sec:A2}}
Consider the product $\text{\boldmath$\Omega$}^{(q)} \cdot \text{\boldmath$\Omega$}^{(r)}$. For $j>k-q-r$,
\begin{multline}
\left( \text{\boldmath$\Omega$}^{(q)} \cdot \text{\boldmath$\Omega$}^{(r)} \right)_{k,j}=\sum_{i=1}^{n} \Omega^{(q)}_{k,i}  \Omega^{(r)}_{i,j}  \\
=\sum_{i=1}^{k-q} \Omega^{(q)}_{k,i} \underbrace{\Omega^{(r)}_{i,j}}_{0\ (\text{since } j>i-r)} + \sum_{i=k-q+1}^{n} \underbrace{\Omega^{(q)}_{k,i}}_{0\ (\text{since } i>k-q)}  \Omega^{(r)}_{i,j} = 0.  \nonumber
\end{multline}

For $j \leq k-q-r$,
\begin{multline}
\left( \text{\boldmath$\Omega$}^{(q)} \cdot \text{\boldmath$\Omega$}^{(r)} \right)_{k,j}
=\sum_{i=1}^{j+r-1} \Omega^{(q)}_{k,i} \underbrace{\Omega^{(r)}_{i,j}}_{0\ (\text{since } j>i-r)}\\
+\sum_{i=j+r}^{k-q} \Omega^{(q)}_{k,i}  \Omega^{(r)}_{i,j}
+\sum_{i=k-q+1}^{n} \underbrace{\Omega^{(q)}_{k,i}}_{0\ (\text{since } i>k-q)}  \Omega^{(r)}_{i,j}.  \nonumber
\end{multline}

Using Eq.~(\ref{eq:omegar}),
\begin{equation}
\left( \text{\boldmath$\Omega$}^{(q)} \cdot \text{\boldmath$\Omega$}^{(r)} \right)_{k,j}
=\sum_{i=j+r}^{k-q} \omega^{(q)}_{k-i-q+1}  \omega^{(r)}_{i-j-r+1}. \nonumber
\end{equation}

Substituting $i=m+j+r-1$,
\begin{equation}
\left( \text{\boldmath$\Omega$}^{(q)} \cdot \text{\boldmath$\Omega$}^{(r)} \right)_{k,j}
=\sum_{m=1}^{k-j-q-r+1} \omega^{(q)}_{k-j-q-r+2-m}  \omega^{(r)}_m. \nonumber
\end{equation}

Thes results about the elements of the product $\text{\boldmath$\Omega$}^{(q)} \cdot \text{\boldmath$\Omega$}^{(r)}$ can be summarized as follows,
\begin{multline} \label{eq:omegaproduct}
\left( \text{\boldmath$\Omega$}^{(q)} \cdot \text{\boldmath$\Omega$}^{(r)} \right)_{k,j}\\
= \left\{ \begin{array}{ll}
            0, &j>k-(q+r), \\
            \sum_{m=1}^{k-j-q-r+1} \omega^{(q)}_{k-j-q-r+2-m}  \omega^{(r)}_m, &\text{otherwise.}
        \end{array}
 \right. 
\end{multline}

Comparing Eq.~(\ref{eq:omegar}) and (\ref{eq:omegaproduct}), we notice that the product $\text{\boldmath$\Omega$}^{(q)} \cdot \text{\boldmath$\Omega$}^{(r)}$ is an omega matrix of degree $q+r$. Thus, using our notation, we can denote the product of $\text{\boldmath$\Omega$}^{(q)}$ and $\text{\boldmath$\Omega$}^{(r)}$ as $\text{\boldmath$\Omega$}^{(q+r)}$. Using Eq.~(\ref{eq:fundvector}), the fundamental vector of this product matrix is given as,
\begin{multline} \label{eq:omegaqplusr}
\omega^{(q+r)}_k=\Omega^{(q+r)}_{n,n-q-r-k+1}=\left( \text{\boldmath$\Omega$}^{(q)} \cdot \text{\boldmath$\Omega$}^{(r)} \right)_{n,n-q-r-k+1},\\
\omega^{(q+r)}_k=\sum_{m=1}^{k} \omega^{(q)}_{k+1-m}  \omega^{(r)}_m, 1 \leq k \leq n-q-r.
\end{multline}

Thus, the fundamental vector of $\text{\boldmath$\Omega$}^{(q+r)}=\text{\boldmath$\Omega$}^{(q)} \cdot \text{\boldmath$\Omega$}^{(r)}$ is obtained through the convolution of the fundamental vectors of $\text{\boldmath$\Omega$}^{(q)}$ and $\text{\boldmath$\Omega$}^{(r)}$ as given in Eq.~(\ref{eq:omegaqplusr}). Let us define the convolution operation $z=\mathrm{Conv}[\textbf{\textit{x}},\textbf{\textit{y}};l]$ between two vectors $\textbf{\textit{x}}$ and $\textbf{\textit{y}}$ to yield another vector $\textbf{\textit{z}}$ of length $l$ as follows,
\begin{multline} \label{eq:conv}
z_k=(\mathrm{Conv}[\textbf{\textit{x}},\textbf{\textit{y}};l])_k\\
\equiv \sum_{m=1}^{k} \textbf{\textit{x}}_{k+1-m}  \textbf{\textit{y}}_m, 1 \leq k \leq l.
\end{multline}

Note that $l$ cannot be greater than the length of any of the input vectors $\textbf{\textit{x}}$ and $\textbf{\textit{y}}$. Here, we have used semicolon to separate $l$ from the two formal inputs of the convolution operation. Moreover, it is straightforward to prove that the convolution operation as defined above is commutative that is $\mathrm{Conv}[\textbf{\textit{x}},\textbf{\textit{y}};l]=\mathrm{Conv}[\textbf{\textit{y}},\textbf{\textit{x}};l]$.
Using the definition provided in Eq.~(\ref{eq:conv}), we can write Eq.~(\ref{eq:omegaqplusr}) as follows,
\begin{equation} \label{eq:omegaqplusrconv}
\text{\boldmath$\omega$}^{(q+r)}=\mathrm{Conv}[\text{\boldmath$\omega$}^{(q)},\text{\boldmath$\omega$}^{(r)};n-q-r].
\end{equation}
From the commutativity of the convolution operation, it also follows that the multiplication of two omega matrices is commutative.

\subsection{Product of an omega matrix and a vector \label{sec:A3}}
Consider a the product of an omega matrix of size $n$ and degree $r$ with an $n \times 1$ vector $\text{\boldmath$\beta$}$. Consider the $k$th element of the resultant vector. For $k \leq r$,
\begin{equation}
\left( \text{\boldmath$\Omega$}^{(r)} \cdot \text{\boldmath$\beta$} \right)_k
=\sum_{m=1}^{n} \underbrace{\Omega^{(r)}_{k,m}}_{0\ (\text{since } m>k-r)}  \beta_m=0. \nonumber
\end{equation}

For $k>r$,
\begin{multline}
\left( \text{\boldmath$\Omega$}^{(r)} \cdot \text{\boldmath$\beta$} \right)_k
=\sum_{m=1}^{n}\Omega^{(r)}_{k,m} \beta_m\\
=\sum_{m=1}^{k-r}\Omega^{(r)}_{k,m} \beta_m
+\sum_{m=k-r+1}^{n} \underbrace{\Omega^{(r)}_{k,m}}_{0\ (\text{since } m>k-r)}  \beta_m,\\
=\sum_{m=1}^{k-r}\omega^{(r)}_{k-r+1-m} \beta_m =(\mathrm{Conv}[\text{\boldmath$\omega$}^{(r)},\text{\boldmath$\beta$};n-r])_{k-r} . \nonumber
\end{multline}

Thus, the non-trivial elements of the vector $\text{\boldmath$\Omega$}^{(r)} \cdot \text{\boldmath$\beta$}$ are given as,
\begin{equation} \label{eq:omegarbeta}
\left( \text{\boldmath$\omega$}^{(r)} \cdot \text{\boldmath$\beta$} \right)_{r+1:n}=\mathrm{Conv}[\text{\boldmath$\omega$}^{(r)},\text{\boldmath$\beta$};n-r].
\end{equation}

For the special case of $r=0$,
\begin{equation} \label{eq:omega0beta}
\text{\boldmath$\omega$}^{(0)} \cdot \text{\boldmath$\beta$}=\mathrm{Conv}[\text{\boldmath$\omega$}^{(0)},\text{\boldmath$\beta$};n].
\end{equation}

\subsection{Advantage of the omega matrices \label{sec:A4}}
We have defined a special class of matrices, the omega matrices, which can be economically represented by their corresponding fundamental vectors. We have shown that for the matrices of this kind we can reduce the operations of matrix multiplications to vector convolutions .


%

\end{document}